# Annealing of Au, Ag and Au-Ag alloy nanoparticle arrays on GaAs (100) and (111)B

*Alexander M. Whiticar[1], Erik K. Mårtensson[2], Jesper Nygård[1], Kimberly A. Dick[2,3\*] and Jessica Bolinsson[1\*]*

[1] Center for Quantum Devices & Nano-Science Center, Niels Bohr Institute, University of Copenhagen, Universitetsparken 5, 2100 Copenhagen, Denmark.

[2] Solid State Physics/NanoLund, [3] Center for Analysis and Synthesis, Lund University, 221 00 Lund, Sweden.

ABSTRACT

Part of developing new strategies for fabrications of nanowire structures involves in many cases the aid of metal nanoparticles (NPs). It is highly beneficial if one can define both diameter and position of the initial NPs and make well-defined nanowire arrays. This sets additional requirement on the NPs with respect to being able to withstand a pre-growth annealing process (i.e. de-oxidation of the III-V semiconductor surface) in an epitaxy system. Recently, it has been demonstrated that Ag may be an alternative to using Au NPs as seeds for particle-seeded nanowire fabrication. This work brings light onto the effect of annealing of Au, Ag and Au-Ag alloy NP arrays in two commonly used epitaxial systems, the Molecular Beam Epitaxy (MBE) and the Metalorganic Vapor Phase Epitaxy (MOVPE). The NP arrays are fabricated with the aid of Electron Beam Lithography on GaAs 100 and 111B wafers and the evolution of the NPs with respect to shape, size and position on the surfaces are studied after annealing using Scanning Electron Microscopy (SEM). We find that while the Au NP arrays are found to be stable when annealed up to 600 °C in a MOVPE system, a diameter and pitch dependent splitting of the particles are seen for annealing in a MBE system. The Ag NP arrays are less stable, with smaller diameters (≤ 50 nm) dissolving during annealing in both epitaxial systems. In general, the mobility of the NPs is observed to differ between the two the GaAs 100 and 111B surfaces. While the initial pattern is found be intact on the GaAs 111B surface for a particular annealing process and particle type, the increased mobility of the NP on the 100 may influence the initial pre-defined positions at higher annealing temperatures. The effect of annealing on Au-Ag alloy NP arrays suggests that these NP can withstand necessary annealing conditions for a complete de-oxidation of GaAs surfaces.

I. INTRODUCTION

Metal nanoparticles (NPs) are interesting from the prospect of using them in a rather broad repertoire of applications as well as for fundamental studies concerning the properties of metals when reducing the volume down to the nanometer scale. Owing to an increased surface-to-volume ratio and their electronic properties, noble metals such as gold (Au) and silver (Ag) exhibit an increased photochemical activity as NPs compared to their bulk form [1]. This has made them interesting for applications involving catalytic chemical transformations [2]. In addition, the size-dependent optical properties of these type of NPs have been shown to be promising for plasmonic nanosensor applications, reaching a detection limit on the single molecule level [3-5]. In materials science, metal NPs are frequently used when synthesizing semiconductor nanowires (NWs) by the particle seeded/catalyzed NW growth approach [6]. In particular, Au NPs have proven to be successful for making III-V semiconductor NWs [7] and Au NP seeded NW growth is one of the main strategies for the directional crystallization of III-V NWs. Commonly, particle seeded NW growth is possible due to the Vapor-Liquid-Solid mechanism (VLS) and the actual NW synthesize/growth process is in general initiated by transforming the NPs into liquid alloy droplets during an annealing step. It has also been shown that solid NPs (Vapor-Solid-Solid, VSS) can aid the crystallization process, making the particle-seeded approach applicable to a rather wide range of growth parameters [6]. Further, Au NPs are not a prerequisite for fabricating NW structures as evident by the often used self-catalyzed method [8, 9] and reports of using other foreign metal NPs [10].

Recently, Ag has emerged as a promising seed material for III-V semiconductor NW growth and may offer some benefits compared to Au [11-14]. Reports also suggest that another option is to engineer alloy NPs composed of two metals [15, 16]. For example, Chou *et al* demonstrated impressive group IV NW heterostructures using Au-Ag alloy catalysts [15]. Indeed, by initially combining two metals into a new type of alloy NP offers a new route to make advanced NW structures as one can engineer the NPs properties to match requirements of, for example, growing by VLS or VSS mechanism, NW growth direction and NW heterostructure junctions by carefully choosing both metal types and alloy composition.

In this work we investigate the effect of annealing on electron beam lithography (EBL) defined Au, Ag and Au-Ag alloy NP arrays [17]. Emphasis is on mimicking conditions during the pre-growth heat treatment included when using metal NPs for NW growth by Molecular Beam Epitaxy (MBE) and Metalorganic Vapor Phase Epitaxy (MOVPE), the two most common epitaxy systems used for the epitaxial growth of III-V materials. We note, however, that our observations are of importance for the general understanding of the evolution of the shape, size and ordering of lithographically defined Au, Ag and Au-Ag NP arrays on III-V semiconductor wafers in general when exposed to heat treatment. This work may therefore be of some importance considering for example applications involving the NPs plasmonic properties as well since size and shape of metal NPs influence their localized surface plasmon resonance [18-20].

It is interesting to note that at present, the influence of the pre-growth annealing step on ordered NP arrays is not very well understood. The main research focus on annealing effects has been devoted to NP formation during so-called thin film annealing [21-26]. However, although the size, density and position of the NPs is pre-defined when using lithography techniques in the NP fabrication process, it is crucial to know if and how these pre-defined properties are effected by an annealing step. With the constantly increasing demand of control in composition and crystal quality of NWs for nano-device applications, well-ordered NW arrays with homogenous dimensions are highly desirable. In general, NP size can be controlled by using for example commercial available colloidal NPs, by using an aerosol method to make and deposit the NPs or by using so-called thin film annealing to some extent [26]. However, all these types of NPs have the drawback of including randomness in the NP deposition across a surface since only the overall or total density of NPs can be defined using these types of NPs. Not being able to have full control of the density and size distribution adds some variations from NW to NW from the same sample for most growth processes. For example, a random distribution of NPs across the substrate surface may influence the local growth rate of individual NWs, resulting in NWs with a length distribution across the sample, as well as variations in the composition within ternary and doped NW structures (when the diffusion length of the involved species differ). In addition, growth rate variations from NW to NW across the sample can be detrimental when fabricating heterostructured NWs since achieving reproducible/homogenous thicknesses of the shell/core and segments within the NWs becomes problematic. The extra level of control reached when being able to also pre-define the position the NWs, and at the same time control both the total and the local density for every NW across the sample, is made possible only when employing a lithography technique [17, 27, 28].

With the increased attention on using lithography techniques for making ordered nanowire arrays [29, 30], we believe that a systematic investigation of EBL defined NPs as the one we report here is crucial in order to increase knowledge and attention about possible effects which are caused by the pre-growth annealing conditions. To the best of our knowledge, this is the first report involving a systematic investigation of the effect of annealing of Au, Ag, and Au-Ag alloy NP arrays. Our study involves a total of about 50 samples/experiments and we use Scanning Electron Microscopy to analyze the NP arrays post-annealing. In this work we explore how different types of NPs (i.e. Au, Ag and Au-Ag alloy) have been affected by annealing in two different environments, the UHV environment of an MBE system and a hydrogen rich flow in a MOVPE reactor - both environments highly relevant for making NW structures. In addition, this study includes annealing experiments at several different temperatures and two different surface orientations that are relevant to growth and processing of III-V semiconductor materials. Focus is on whether or not the pre-defined positioning is preserved and possible shape- and size changes that may occur. Our study also include different initial NP densities (i.e. pitch distances) and diameters to be able to generalize our observations with respect to these two key parameters for ordered NP arrays.

## II. EXPERIMENTAL METHODS AND DESIGN

### II. A. PARTICLE FABRICATION AND DEPOSITION

The metal NPs were defined using Electron Beam Lithography (EBL). First, two layers of electron beam (e-beam) sensitive resist were spin coated onto epi-ready GaAs wafers with either an 111B or a 100 surface orientation resulting in 100 nm of 6% co-polymer and 50 nm of 2% poly-methyl-methacrylate (PMMA). After the deposition of each layer of resist, the wafers were baked on a hotplate (185 °C for 1.5 min). An Elionix ELS-7000 100 keV EBL system was used to define ordered arrays of well-defined circles across the wafers with diameters ranging from 25 to 150 nm and pitch distances ranging from 0.5 to 5 µm as shown in Fig. 1. For the e-beam exposures, an e-beam with a spot size of 5 nm was used and NPs were defined by multi-shot exposures to achieve uniform disks. An area dose of 7600 $\mu$C/cm$^2$ was used to define NPs with a 25 nm diameter and 1000 $\mu$C/cm$^2$ for 100 to 150 nm NPs. To develop the patterns the wafers were submerged into a 1:3 mixture of methyl-isobutyl-ketone (MIBK) and isopropanol (IPA) for 1.5 min, and thereafter rinsed with IPA and ultra clean water. A N$_2$ gun was used to quickly blow-dry the wafers. Before metal deposition, the wafers were treated with oxygen plasma with a rate and a time corresponding to removal of 10 nm of PMMA resist and a short (10 s) dip in 10 % of buffered hydrofluoric acid (bHF). After the bHF treatment, the wafers were transferred in air and quickly loaded into an e-beam evaporation system kept at a base pressure of 2·10$^{-8}$ Torr. The metal film deposition onto the wafers was monitored using a Quartz thickness monitor to give a total film thickness of 15 nm on all wafers. For the Au-Ag alloys, a 7.5 nm thick Au film was deposited first and then directly followed by the deposition of 7.5 nm of Ag. Lift-off was carried out overnight in room temperature Acetone (it was noted that lift-off in 60 °C NMP caused an uncontrolled dendrite oxidation of the Ag NPs). In general, the metal deposition was carried out maximum 2 days before the annealing experiments and stored in a nitrogen environment to minimize the NPs exposure to air [31, 32]. Before loading the samples into the MBE/ MOVPE system, a 40 nm oxygen plasma treatment was done to ensure the removal of any form of residual resist and a 10 s etch in 5 % unbuffered HF to remove the oxide formed during the oxygen plasma treatment.

## II. B. ANNEALING PROCEDURE

Annealing experiments were performed in an MBE or an MOVPE system in the temperature range of 400 °C to 600 °C for the Au and Ag NP arrays and at 600 °C for the Au-Ag alloy NP arrays. In addition, one other annealing temperature was included in the MBE annealing series and one for the MOVPE annealing series for the Ag and the Au NP arrays. For the MBE experiments, all samples were degassed at 250 °C for 1 h in a UHV environment prior to the annealing experiment. This is a standard degassing procedure for MBE systems to protect the UHV environment when introducing new wafers/samples from ambient air. To be able to account for this different degassing procedure compared to MOVPE, we also included this type of annealing into the analysis of the annealing experiments carried out inside the MBE. Further, for the annealing series carried out in MOVPE we also included a 650 °C annealing (i.e. 50 °C higher than the temperature range used in MBE). A background pressure/flow of $As_2$ (MBE) and $AsH_3$ (MOVPE) was used at all times when the samples were exposed to a temperature of 300 °C and above.

For the MBE annealing, a solid source system called Varian GEN II MBE with a base pressure of $10^{-11}$ Torr was used. An $As_2$ flux corresponding to a beam equivalent pressure of $6·10^{-6}$ Torr was turned on when reaching a temperature of 300 °C, the temperature was then increased to desired annealing temperature in about 15 min (wafer temperature was monitored by a pyrometer) and then directly cooled down while keeping the $As_2$ flux until reaching below 300 °C.

For the MOVPE annealing, we used an AIXTRON Close Coupled Showerhead (CSS) 3x2" MOVPE reactor. The base pressure of the reactor is 75 Torr (100 mbar) with a total gas flow of 8 slm, using hydrogen as a carrier gas. Prior to the experiments, the reactor chamber was covered with a buffer-layer of GaAs, to ensure a clean environment for GaAs substrates. For the annealing experiments, the temperature was increased at a rate of 1 °C/s until target temperature was reached. A background flow of arsine was turned on when the sample was above 300 °C, corresponding to an AsH3 molar fraction of $2.5 \cdot 10^{-3}$. The annealing conditions were maintained for 7 min after which the samples were cooled down. The temperature of the MOVPE reactor is measured using a thermocouple, which has been calibrated using in-situ optical spectroscopy of the reactor. The sample temperature is expected to slightly below the set temperature, but no more than 30 °C according to in-situ measurements of reference wafers using optical reflectometry. Characterization of the annealing experiments was performed by Scanning Electron Microscopy (SEM) using a JEOL 7800F equipped with an Oxford Instruments energy-dispersive X-ray spectroscopy detector (EDS) and a Raith eLine 100 system and acceleration voltages in the range between 1 to 20 keV.

III. RESULTS AND DISCUSSION

Results from this work is presented and discussed one particle type at the time in separate sub sections below, including possible similarities and difference that we found between the two epitaxy systems. For most of the figures and in the main part of the results, we will focus on the 100 nm NP diameter arrays with 0.5 $\mu$m pitch distance as a standard to describe general trends for the different types of NPs and comparison between them. However, as described in section II. A, other pitch distances and NP diameters were also included. Observations regarding influence of annealing with respect to diameter and pitch distance, when this differ from our standard as defined above, will be pointed out clearly below and is illustrated in separate figures either in this section or in the Supplemental Information file associated with this letter.

III. A. ANNEALING OF AU NANOPARTICLES ON GAAS 111B AND 100 SURFACES

In this section, we first summarize our observations concerning the annealing of the Au NPs in the MBE system. After this, we will move on to what we observed when changing to a MOVPE system and highlight the similarities and differences between annealing of Au NP arrays inside the two different types of epitaxy systems.

An overview of the evolution of the Au NP arrays for the full temperature range that we studied using the MBE system is shown in Fig. 2. At the lowest temperature, 250 °C, the Au NPs have similar diameter to height ratio as the initial Au disks shown in Fig. 1 (the initial diameter and height of the Au disks of the ones shown here were defined by the EBL process and film thickness to be 100 nm and 15 nm respectively). However, the Au NPs morphology have changed from the initial circular shape (see Fig. 1d), indicating that the NPs are somewhat affected by the 250 °C annealing. As seen from Figure 2 a) and e), the morphology of the Au NPs after the 250 °C annealing depend on the substrate orientation: The Au NPs on the (111)B surface are observed to have a triangular shape (a) while on the (100) the Au NPs are adapting a more rectangular shape (e). As mentioned in section II. B above, an annealing at this temperature mimic the initial degassing step commonly used when inserting wafers/substrates into the UHV system that contain the MBE machine. The difference between the two surface orientations suggest two things: 1) the metal atoms are already mobile at this low temperature and 2) the surface orientation plays a role for the re-shaping of the Au NPs. The actual morphology difference between the Au NPs on the (111)B and the (100) surface is not surprising considering the difference in symmetry between the two. The 111B surface of a zincblende crystal have a 3-fold symmetry and the 100 surface have 4-fold, which suggest that the interfacial energy between the Au NPs and the substrate surface strongly influence the morphology of the Au NPs after the annealing. Further, we argue that surface oriented re-shaping that is observed also indicate that the Au NPs are epitaxial attached to the substrate surface after the annealing. Since the total volume of the Au NPs is not significantly changed, the alloying between the initial Au NPs and the substrate material is probably rather limited and we believe that the Au NPs have remained solid during the annealing at this temperature. The melting temperature for Au is much higher (above 1000 °C for pure Au) and reports in the literature on the Au-GaAs system suggest that the transformation from solid Au into a liquid Au-Ga alloy on GaAs (111)B occurs at a temperature which is about 100 °C higher [33, 34].

When increasing the annealing temperature to 400 °C, the shape of the Au NPs changes dramatically as seen in the inset in Fig. 2b). The volume is increased significantly, indicating an alloying between the Au and the GaAs substrate and possibly a transition from solid NP to liquid droplets (LDs) [33][35, 36]. This is in agreement with earlier report of an increased uptake of Ga in the Au NPs/LDs at this temperature [33, 34].

At 500 °C and above, triangular and rectangular shaped pits have formed in the vicinity of the Au NPs on the GaAs (111)B and (100) surfaces, respectively. The formation of pits in the surfaces is likely caused by local decomposition of the substrate in the proximity of the Au LDs [37] and the high solubility of Ga in Au in this temperature range [35, 36, 38]. EDS point measurement on cross-sectional samples confirmed an increased Ga content and a shift in the Ga:As stoichiometry towards more Ga rich when approaching the Au NPs/LDs on the surface (see Supplemental Information I) [39]. Since the congruent evaporation temperature of GaAs has been found to be controlled by the flux of the arsenic species [40], we expect that a change in $As_2$ flux at temperatures above ~ 500 °C can influence the size and depth of pit formation. It is possible that an increased As flux would make it possible to avoid pits while a decreased flux would lead to more sever pit formation or that these will start to form already at lower temperatures than what we report here. Looking at the Au-Ga-As ternary and Au-Ga binary phase diagrams data [35, 36] we note that although the solubility of As in Au is very low, the presence of As still influence the solubility of Ga in Au and the melting point of the GaAs+Au mixture [35, 36].

For a 600 °C annealing, the rectangular pits on the (100) become larger/longer and spread out along the <110> directions, evolving into trenches (see Fig. 2 h)). In conjunction with the formation of these trenches, we also observe that the Au NPs slightly deviates from their initial position in the <110> directions (i.e. inside the trenches). As mentioned earlier, our study includes several different diameters and pitch distances (see Fig. 1). When turning to other diameters and pitch distances for the 600 °C annealing we note that in addition to the moving along the <110> directions, the Au NPs also splits up for the (100) surface for certain diameters and pitch distances. In Fig. 3 we illustrate this splitting versus diameter and pitch distance by showing what happens when the particle diameter is increasing from 25 nm to 150 nm (diameter is increasing going from left to right) for the pitch distances of 0.5 $\mu$m (upper row, a-d) and 1 $\mu$m (lower row, e-h), respectively. From Fig. 3 a-d it can be seen that as the diameter of the Au NP increases, the splitting behavior decreases. For 0.5 $\mu$m spacing, no splitting is observed for the 150 nm (d) and the 100 nm (c) NP diameters, while for 50 nm (b) and below the Au NPs split up inside the trenches. Notable is that although the 100 nm Au NPs do not appear to split up for the smaller pitch distance (0.5 $\mu$m, see Fig. 3c), for a larger pitch distance it does (1 $\mu$m, see Fig. 3g).

It is not obvious why there is a pitch distance and diameter dependence of the Au NP splitting. When increasing the pitch or decreasing the diameter, the total amount of Au on the surface decreases. This would imply a decreased decomposition of the substrate surface with increased pitch distance and decreasing Au NP diameter (and vice versa) if the Au somehow support this process as discussed above [37]. Thus, the splitting is more pronounced at conditions for which the decomposition of the substrate should be less (i.e. larger pitch distances and smaller Au NP diameters) – leading up to ask the question if it is related to a lack of available Ga species on the surface. However, increasing the amount of Au by having a larger density of Au NPs, or by having larger diameters, is also increasing the uptake of available Ga when forming more or larger Au-Ga alloy LDs. It is therefore not clear whether the splitting occurs for conditions where there is more or less Ga available. Nevertheless, it can be concluded that at a certain upper limit of the density of NPs, the decomposition "zones" for the different NPs will overlap and thus lead to less material available per NP compared to for lower densities.

Recently, Zakharov et al [37] investigated the dynamics of gallium droplets on GaP (111)B during annealing in vacuum (no arsenic flux/flow during the annealing). Among other things, they showed how $\mu$m-sized Ga droplets may stretch out into an elongated form corresponding to as much as twice their own original diameter across the surface within less than 30 s, and then abruptly retract back to a more symmetric shape at, or close to, its original position. In their study, this phenomenon was observed for droplets where the mobility was reduced by lowering the annealing temperature while at higher temperatures the droplets were found to move around much more on the surface. By comparing surface areas with and without Au NPs, they noticed a clear difference: In the areas where the Au is present, the de-oxidation and other surface transition occurred much faster and the moving droplets were found to deflect from these areas. In our case, the Au NPs/LDs are distributed in a periodic manner across the surface. Overall, a higher density (i.e. smaller pitch distance) or larger Au NP diameter could therefore effectively work to limit the overall movement of the Au LDs, assuming that the Au-Ga alloy LDs behave somewhat similar to the gallium droplets discussed by Zakharov et al [37].

The splitting that we observe for smaller diameters and larger pitch distances shown in Fig. 3 could thus be explained by a higher overall mobility of the Au-Ga alloy LDs in these arrays. Considering the stretch/retract dynamics mentioned above, this may very well cause the LDs to split up as they moved around at elevated temperatures during the annealing or during the cooling down step. It should be noted that the annealing and cooling down process in this work was carried out to resemble realistic annealing conditions commonly used prior to initiating NW growth in the two epitaxy systems, and not with respect to ensure an equilibrium state of the NPs at the different temperatures.

It is interesting to note that the movement observed for so called running gallium droplets on GaAs (100) is reported to occur preferentially along two of the <110> directions of the surface [41], which correlates with the trenches we observe here and thus further strengthen the that these are related to movement of LDs during the annealing. Further, we believe interpretation that the main reason for why the splitting and movement of the Au-Ga alloy LDs is seen for the (100) surface and not for the (111)B is due to differences in surface energies between the two systems, in particular the interfacial energy between the Au-Ga alloy droplet and the substrate surface [41]. The anisotropic behavior of the trench formation on the (100) is most likely related to interaction between the Au-Ga LDs and the anisotropy of the (100) surface, where the later introduce an energy barrier difference depending on if the LDs move/stretches in paths that are parallel or perpendicular to the As dimers [42, 43].

We now turn to annealing of the Au NPs in a MOVPE system. Compared to what have been presented above for the MBE system, we find that the Au NPs are in general more stable during annealing in a MOVPE system with respect to maintaining their initial pre-defined positions. In Fig. 4 we show an overview of the evolution of the Au NPs and the surfaces during an annealing in the MOVPE system. Here we included also an annealing of 650 °C, which is slightly above the commonly used annealing temperature used in the MOVPE system before nanowire growth. In addition, if there is a temperature offset between the two systems (or between the set temperature and sample temperature for the MOVPE experiments, see section II. B) we also ensure that we reach at least as high temperature in the MOVPE system as for the MBE one. For the MOVPE annealing experiments, we do not observe any splitting of the Au NPs irrespectively of the Au NP diameter and pitch distance within the Au NP arrays. (It should be noted that the halo around the Au NPs that is seen in Fig. 4e and f are formed during the processing of the NPs and is not a result of the annealing.) Similar to what is observed for annealing in the MBE system; the Au disks are transformed to resemble half spheres already at 400 °C (see inset in Fig. 4a). However, at 600 °C, instead of forming voids in the surface around the Au NPs as observed for the annealing in the MBE, the particles begin to creep/move while forming tail-like/planar NW structures along the substrate surface. For a higher annealing temperature of 650 °C, the Au NPs become more mobile and are sometimes found to creep rather far distances on the (100) surface in particular (see Fig. 4h). As seen, for the (100) surface these "tails" or NWs form in the same directions as the trenches were formed for the MBE annealing – in two of the <110> directions. Further, the "diameter" of the planar NWs also appears to be more or less the same as the Au NPs. Interestingly, the size of the Au NP/LD is observed to increase as they move a distance that is further than the pitch distance. Looking at Fig. 4h) it is seen that the Au NPs/LDs seems to "collect" other Au NPs/LDs that they find along their path, and become larger by the merging of multiple NPs. Note also that this results in the "tail" to become wider. The length of the planar NW is observed to increase with Au NP diameter (see Supplemental Information II) and comparing the two surface orientations reveals that the planar NWs grow much longer for the (100) surface (see Fig. 4d/h), while no discernible pitch dependence is found for either surface.

From what is outlined above, we can conclude that while trenches/voids are formed as the Au NPs/LDs move when annealing them in the MBE system, some kind of planar NWs are formed in the MOVPE system – suggesting a fundamental difference between the two systems. It should be noted, that growth of planar NWs along the substrate surface is a sub-field within NW growth [44-47]. However, growth of NWs in general implies a continuous supply of group III as well as group V. Here we have no intentional source of group III material on during our annealing experiments. It is clear, that the change from trenches to planar structures between the two systems must mean that either the condition to reach supersaturation or equilibrium of Ga in the Au NPs/LDs is higher in the MBE system compared to the MOVPE system, or that there is a higher unintentional background of group III material somehow in the later. We note that there is probably a difference in the de-oxidation temperature and process between the two systems. Since the carrier gas in the MOVPE is hydrogen, which is known to support de-oxidation of III-V semiconductor surfaces [48], the de-oxidation process is probably more efficient here compared to the UHV environment inside the MBE system. At present, we cannot say whether this is a contributing factor or not to the differences which we observe.

III. B. ANNEALING OF AG NANOPARTICLES ON GAAS 111B AND 100 SURFACES

Similar to the section above regarding the Au NPs, we will present the observations concerning the MBE system first. However, we will also include comparisons with the observations presented above of the annealing of the Au NPs in the same system. After this, we will present our findings for the annealing of the Ag NPs in the MOVPE system and compare the effect of annealing of Ag NPs by the two different epitaxy systems (i.e. following more or less a similar structure as for section III A).

In Fig. 5 we show an overview of the morphology and arrangement of the Ag NPs after annealing in the MBE system (surface in the upper and lower rows are GaAs (111)B and (100), respectively). As noticed by comparing Fig. 5a/5e with Fig. 2a/2e for the 250 °C annealing in MBE, there is a clear difference seen from the SEM images of the morphology of the Ag NPs (Fig. 5) and the Au NPs (Fig. 2): The Ag NPs have transformed into a more droplet-like shape with an increased(decreased) height(diameter) compared to the initial Ag disks. In addition, a few of the Ag NPs have split into two or more NPs. It should be noted that the shape and size of the initial Au and Ag NPs are the same before being exposed to any annealing, so any difference noticed after the annealing process is related to the annealing process itself and not a difference in the initial NPs.

When the annealing temperature is increased to 400 °C, all the Ag NPs splits up into one larger NP and several smaller NPs in its close proximity (see Fig. 5b and f). Although splitting occurs, the overall pre-defined position of the Ag NPs is preserved. The splitting of the Ag NPs is not observed to depend on initial Ag NP diameter or pitch distance of the arrays. The evolution of the Ag NP morphology with annealing temperature can be seen from the insets in Fig. 5. Note in particular the increased wetting angle between the insets shown in a and b, indicating that the NPs have been liquid at some point during the annealing process at 400 °C. We note that the shape of the Ag NPs is more drop-like compared to the Au NPs when annealed to the same temperature and same epitaxy system (see Fig. 2 and Fig. 5). Pure Ag have a melting point about 100 °C lower than pure Au, still far above the annealing temperatures used in this work (melting point of pure Ag is 962 °C [49]. However, the first liquid phase for the Ag-As-Ga ternary system lies just below 300 °C [49] and involves a concentration of Ga of only around 12 %. For the Au-Ga-As system, the liquid phase is reported at higher temperatures and for almost twice as high Ga content (at 400 °C the Au

NPs should contain about 22 % of Ga to be liquid [35, 36]. One would thus expect that solid to liquid phase transition occur at lower temperature for Ag compared to Au.

After a 500 °C annealing the splitting is enhanced (see Fig. 5c and g), and there are many small Ag NPs at the initial NP positions (on average about 6-7 per position in the array) with diameters ranging from about 5 to 25 nm. At this annealing temperature, we also note a size dependent effect: The smaller Ag NPs appear to dig pits into the surface while the larger ones do not. This is seen both for the smaller Ag NPs that have been formed by shedding from the larger ones and for the arrays with the smallest initial NP diameter (i.e. the ones pre-defined by the EBL process to be 25 nm in diameter). We note that for the size dependent effects also occur for the Au NPs, where the diameter dependent mobility and splitting illustrated in Fig. 3 is also observed after a 500 °C annealing for 25 nm Au NPs. However, while the smaller Ag NPs appear to dig into the substrate the Au NPs move and split up along the trenches in the surface, indicating a fundamental difference between the two. Also, for the Ag NPs the splitting effect does not appear to be dependent on NP diameter, pitch distance or surface orientation, while for the Au NPs the splitting and increased mobility is observed for the (100) surface only. A comparison between the type of splitting and mobility observed for the Au and the Ag NPs is illustrated in Fig. S3 in the Supplemental Information.

For the highest annealing temperature investigated in the MBE system, 600 °C, the number of NPs decreases and larger NPs are found at or close to the initial pre-defined Ag NP position within the arrays (see Fig. 5d and h). At this temperature, the 25 nm NP arrays have entirely vanished from the surface, which reveals difficulties in retaining patterns for smaller Ag NPs (see Supplementary Information IV). Slight deviations in NP diameters are observed for the larger diameters as seen in Fig. 5d (i.e. the NP diameters are smaller compared to the initial diameters), which we assume is related to smaller part of the Ag NPs being shedded off at some point during the annealing processes and then disappearing from the surface in a similar manner as for the smaller initial diameters mentioned above. At this annealing temperature we also note a difference between the two surfaces: For the (111)B the Ag NPs is more or less positioned at their initial positions and etched triangular pits similar to the Au NPs, while the Ag NPs appeared to have moved around more on the (100) surface. For the later, the Ag NPs have formed trenches in a similar manner as the Au NPs did (Fig. 2h) and the same type of anisotropy is observed.

In Fig. 6 SEM images of two different diameters (100 nm and 50 nm) and pitches (0.5 $\mu$m and 1 $\mu$m) are shown for the Ag NPs. As seen from Fig. 6, the 100 nm Ag NPs (b) appear to be more mobile and form longer trenches than the 50 nm ones (a). Interestingly, when decreasing the pitch distance for the 100 nm Ag NPs, the mobility and trench formation appear to be more suppressed for a larger pitch distance (compare 0.5 $\mu$m (b) and 1 $\mu$m (d) for 100 nm NPs in Fig. 6). We note that both the dependence of pitch and diameter for the Ag NPs with respect to trench formation and mobility on the GaAs (100) surface is thus the opposite of what we observe for the Au NPs. As mentioned in III. A, the tendency for the Au NPs to split up and move from their initial position is dependent on pitch distance and diameter, where larger diameters as well as smaller pitch distances suppressed mobility and splitting of the Au NPs on the GaAs (100) surface.

Now we turn to the annealing of the Ag NPs in the MOVPE system. In Fig. 7 a general overview of the evolution of the Ag NP arrays with respect to annealing temperature is presented. As is seen, the Ag NPs show the same tendency to split up as when being annealed in the MBE system. In fact, the similarity is rather striking. However, when increasing the annealing temperature to 500 °C there is a sudden change when the annealing is done in the MOVPE compared to the MBE. While an annealing temperature of around 600 °C is necessary for annealing in MBE in order have single NPs at each position in the array (instead of many smaller ones), a 500 °C annealing is sufficient to have a significant less shedding of the NPs for annealing carried out in the MOVPE system. In the later, the average number of NPs reduce from 3.5 number of NPs after 400 °C annealing to around 1.5 NPs after annealing at 500 °C and 600 °C at each initial NP position. For the 650 °C MOVPE annealing shown in Fig. 7 d and h, an overall decrease in diameter is seen and the NPs are missing from the array, with no trenches indicating that they moved across the surface. Although this temperature was not included for the MBE annealing we have noted earlier that the Ag NPs appear to disappear from the surface at temperatures higher than 600°C in the MBE system [50]. When attempting to grow Ag seeded GaAs NW at temperatures slightly higher than 600°C, we have repeatedly found a flat mirror-like surface morphology after the growth process, and noted a seemingly absence of both Ag NPs and Ag seeded GaAs NWs on these type of samples [50]. Since the Ag seeded GaAs NWs grow nicely at 600 °C [11], it might be that the Ag NPs themselves becomes unstable at higher temperatures. At present, we cannot conclude whether the Ag from the NPs have been distributed into smaller cluster, or perhaps even single atoms, across the surface and becoming too small to be visualize by SEM imaging or if the Ag material have diffused into the GaAs substrate itself.

In a similar manner as for the Au NPs, the tendency to introduce pits and voids in the surface locally around the Ag NPs is seen using the MBE system but not for the MOVPE system. It is important to point out, however, that the smaller Ag NPs are unstable at the annealing conditions in both systems.

III. C. ANNEALING OF AU-AG NANOPARTICLES ON GAAS 111B AND 100 SURFACES

After exploring the annealing effect on Au and Ag NP arrays, it is interesting to also include a combination of these two metals, Au-Ag alloy NPs, and compare with what is presented above. As described in the experimental section above (II. B Annealing procedure), we used an annealing temperature of 600 °C for this type of NPs. This temperature is the most relevant one, considering it is slightly above the temperature needed for a full de-oxidation of all III-V semiconductor surfaces and an annealing at this temperature is in some cases absolutely necessary in order to achieve excellent optical quality of the final NW structures [51]. Here we present the general observations made for the Au-Ag NPs and compare with the other two types of NPs.

Fig. 8 show SEM images of the Au-Ag NP arrays after annealing at 600 °C in the MBE (a,c) and the MOVPE (b,d) system. Overall, the Au-Ag NPs are rather stable when annealed on the (111)B surface and behave similar to the Au NPs: the pre-defined position is intact after annealing and no sign of shedding as noted for the Ag NPs. When annealed in MBE, the Au-Ag alloy NPs influence the surface in a similar manner as the Au NPs by creating small triangular pits in their vicinity on the (111B) surface (Fig. 8a) while trench-like voids are found on the (100) surface (Fig. 8c). Further, for the (100) surface and the MBE annealing we note that smaller diameters show tendency of splitting (similar to the Au NPs) while larger diameters appear to not to split as much but instead become more mobile. The effect of a 600 °C annealing in the MBE system with respect to Au-Ag NP diameter is shown in Fig. S5 in the Supplemental Information, illustrating that unlike the smaller Ag NPs, the smaller Au-Ag NP arrays can withstand an 600 °C annealing. It is interesting to note that when annealed in the MOVPE system, the Au-Ag NPs are indeed stable for both surface orientations (Fig. 8b and d) and there appear not to be any signs of tail-like/planar NWs starting to form as in the case for the Au NPs (see Fig. 4c and h).

## IV. CONCLUSION

Au NP arrays display a surface orientation dependent influence of annealing already at 250 °C (MBE system, UHV conditions), where the NPs are found to reshape to triangular or rectangular shape on the 111B and the 100 surface, respectively. At 400 °C, the Au NPs show signs of significant alloying with the substrate in both epi-systems and it is likely that they have been transformed into LDs at some point during the annealing. In the MOVPE system, the Au NPs are stable and the patterns are intact for annealing conditions up to around 600 °C. However, going higher in annealing temperature (i.e. 650 °C) is problematic due to an increased mobility of the Au NPs on both surface orientations. Overall, the mobility of the Au NPs is found to be higher on the 100 than the 111B surface, as evident from observations made for both epi-systems. For the MBE system, the difference between the two surface orientations is noticeable already at 600 °C. While the Au NP arrays are observed to be stable for the 111B surface for a 600 °C annealing, this annealing temperature is enough to increased mobility of these NPs in the MBE system and cause a detrimental diameter and pitch dependent splitting within the NP arrays.

In agreement with phase diagram data for the Au-Ga-As and Ag-Ga-As systems, for the Ag NPs the transition to LDs appear to occur at a lower temperature compared to the Au NPs. Further, smaller Ag NPs becomes unstable at too high annealing temperatures in both systems. For both systems (less severe for the MOVPE system compared to the MBE system), some splitting of the Ag NPs is found for arrays being annealed around 400-500 °C. Fortunately, this splitting is not found using a 600 °C annealing temperature. However, for both systems we note a small decrease in particle diameter after annealing at 600 °C, indicating that some Ag material have dissociated during the annealing process. Similar to the Au NP arrays, the Ag NPs are more mobile on the 100 surface compared to the 111B but display an opposite diameter and pitch dependence with respect to mobility compared to what we find for the Au NPs.

Finally, annealing of Au-Ag alloy NP arrays above the de-oxidation temperature for III-V semiconductor materials shows that this type of NP arrays withstand necessary annealing conditions for NW growth process in both MOVPE and MBE systems.

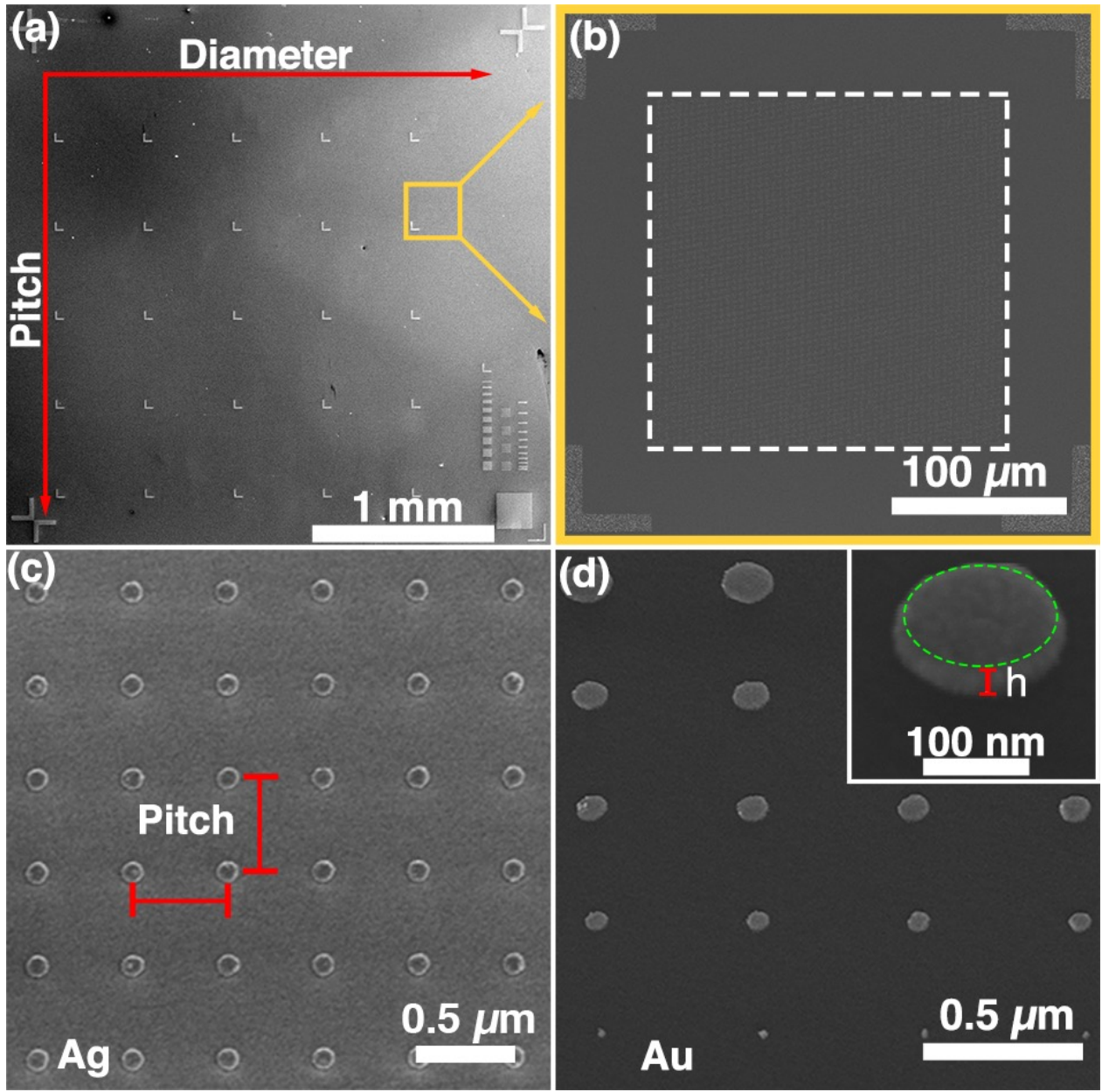

**Figure 1**. Top-view SEM images detailing an overview of the NP arrays used for our annealing experiments. (a) Total lithography design showing the 25 arrays with varying pitch and diameter with (b) showing an overview of an individual array of 200 x 200 $\mu$m in size. (c) 100 nm Ag NPs after lift-off and before annealing. (d) 45° tilted SEM image of Au NPs showing the NP diameters used for annealing ranging from 25 nm up to 150 nm. The inset in (d) shows a 150 nm Au NP with h denoting the deposition thickness of 15 nm.

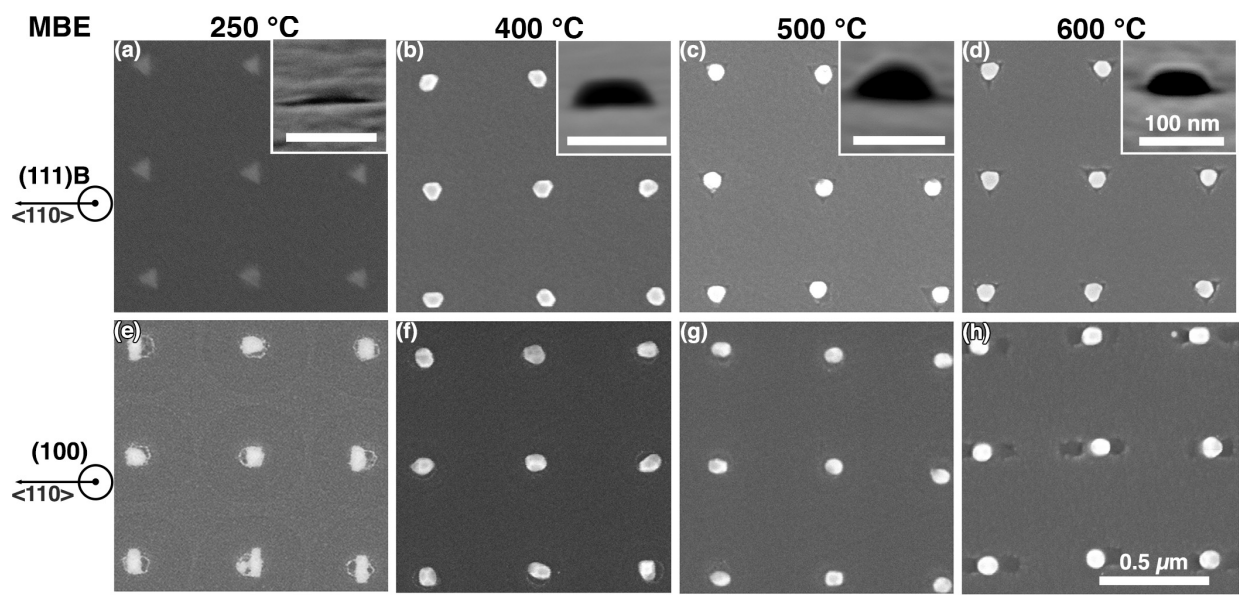

**Figure 2.** Top-view SEM images of MBE annealing of 100 nm Au NPs arrays with a 0.5 µm pitch on GaAs (111)B and (100) substrates with annealing temperatures of 250 °C to 600 °C. All images have the same scale besides the insets that show 85° tilted side-view images of characteristic NPs with 100 nm scale bars.

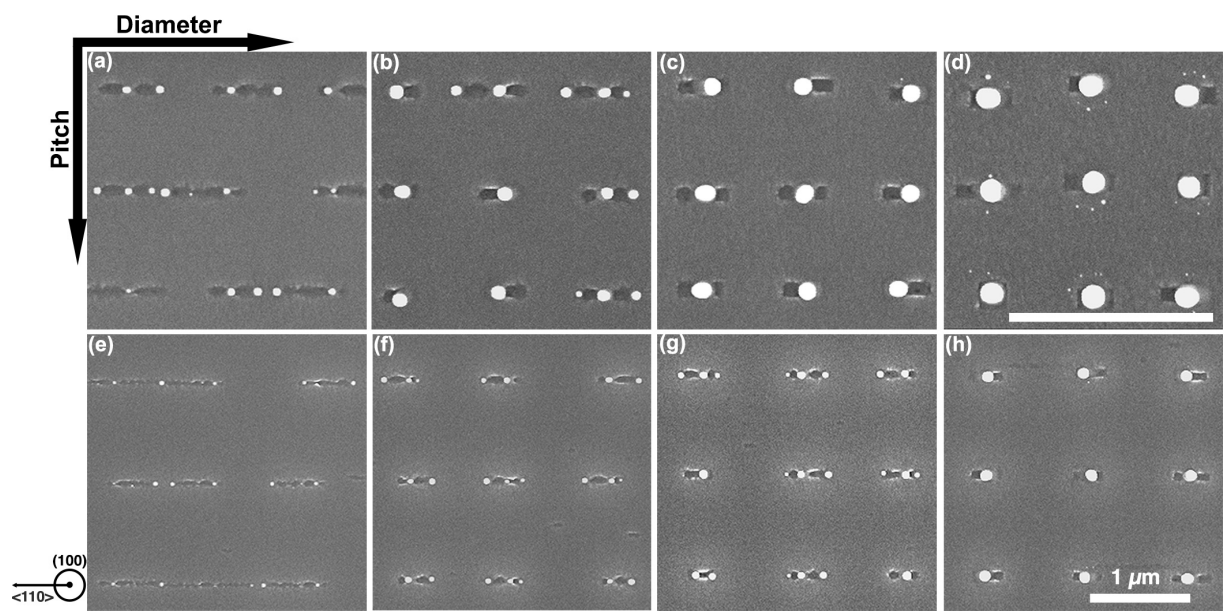

**Figure 3.** SEM images of annealing of Au NPs on GaAs (100) in MBE featuring the diameter and pitch dependence on the splitting behavior of the Au NPs. The arrays

have diameters of (a,e) 25 nm, (b,f) 50 nm, (c,h) 100 nm, (d,h) 150 nm and pitches of (a-d) 0.5 µm and (e-h) 1 µm. (a-d) and (e-h) are the same scales with scale bars of 1 µm.

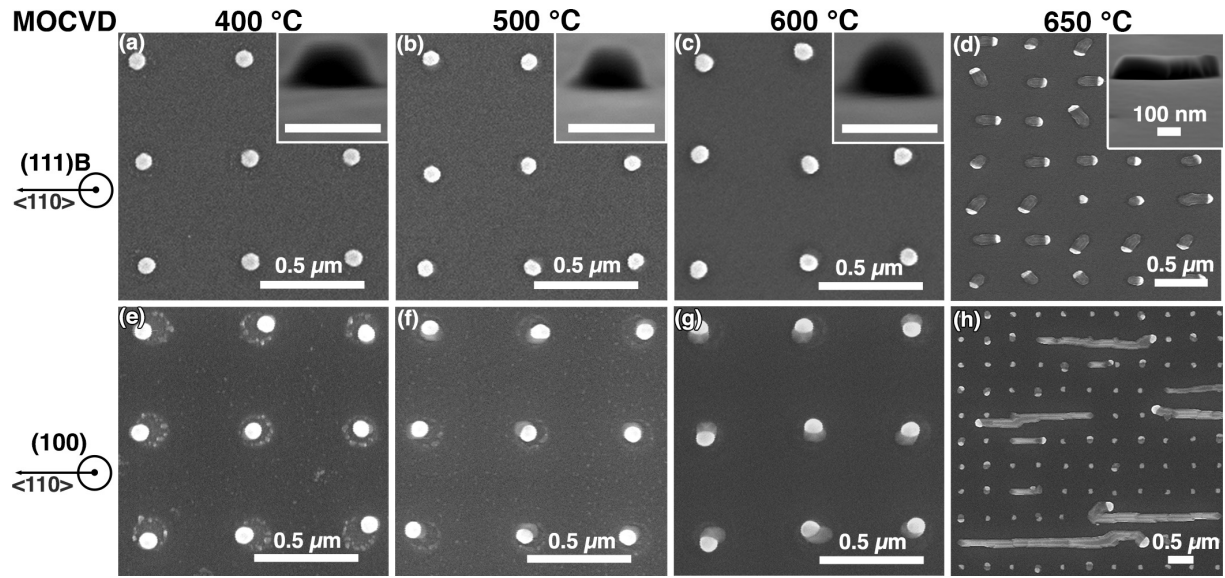

**Figure 4.** SEM images of MOVPE annealing of 100 nm Au NPs arrays with a 0.5 µm pitch on GaAs (111)B and (100) substrates with annealing temperatures of 400 °C to 650 °C. Insets show 85° tilt-view images of characteristic NPs with 100 nm scale bar.

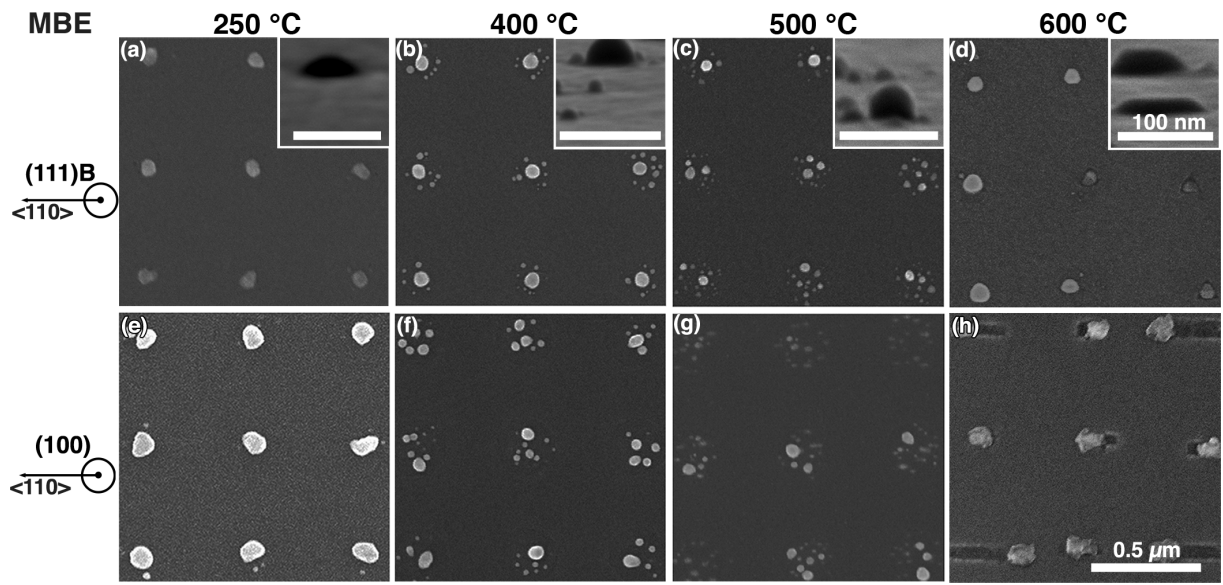

**Figure 5.** SEM images of MBE annealing of 100 nm Ag NPs arrays with a 0.5 µm pitch on GaAs (111)B and (100) substrates with annealing temperatures of 250 °C to 600 °C. All images share the same 0.5 µm scale bar. The insets show 85° side-view images of characteristic NPs with 100 nm scale bars.

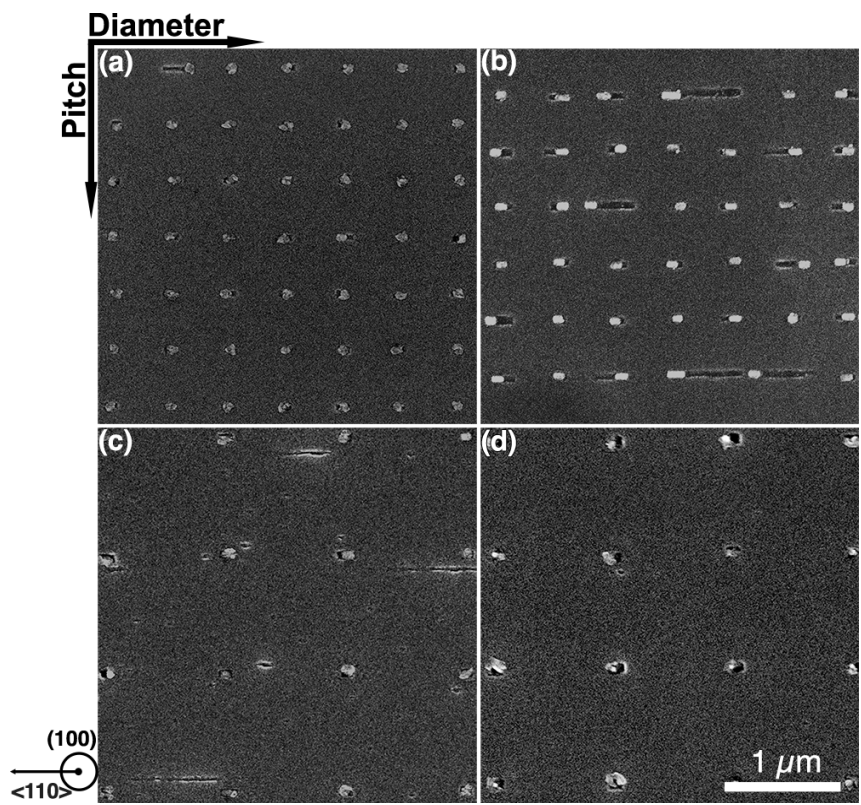

**Figure 6.** SEM images depicting the diameter and pitch dependence for the migratory behavior exhibited by Ag NPs on GaAs (100) for a 600 °C annealing in MBE.

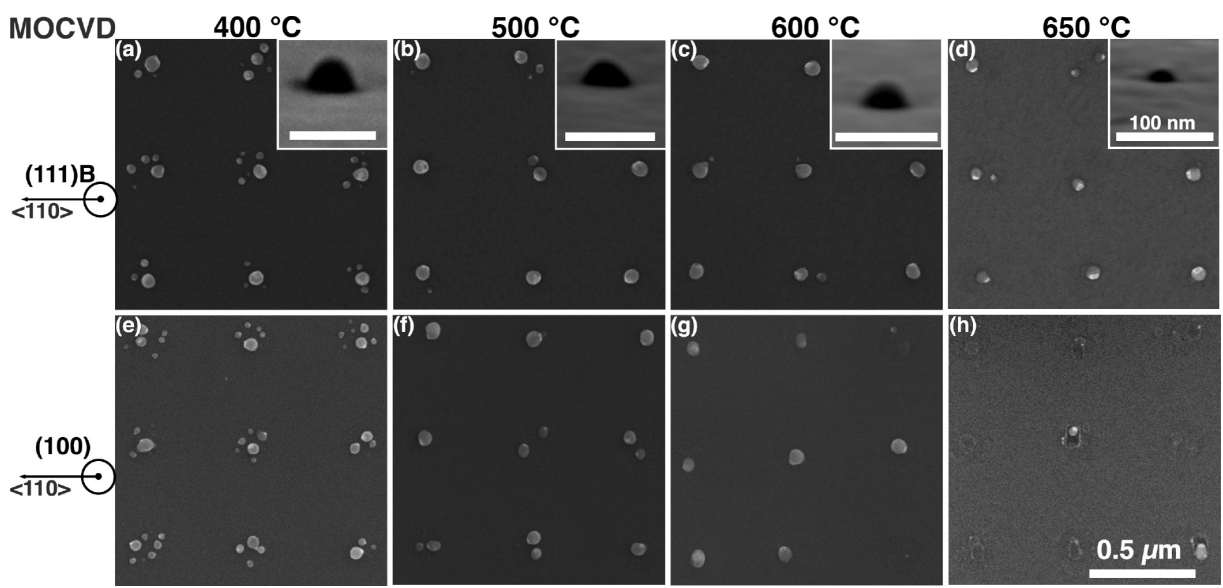

**Figure 7.** Top-view SEM images showing MOCVD annealing of 100 nm Ag NPs arrays with a 0.5 µm pitch on GaAs (111)B and (100) substrates with annealing temperatures of 400 °C to 650 °C. All images share the same 0.5 µm scale bar besides the insets that are 85° side-view images of characteristic NPs with 100 nm scale bars.

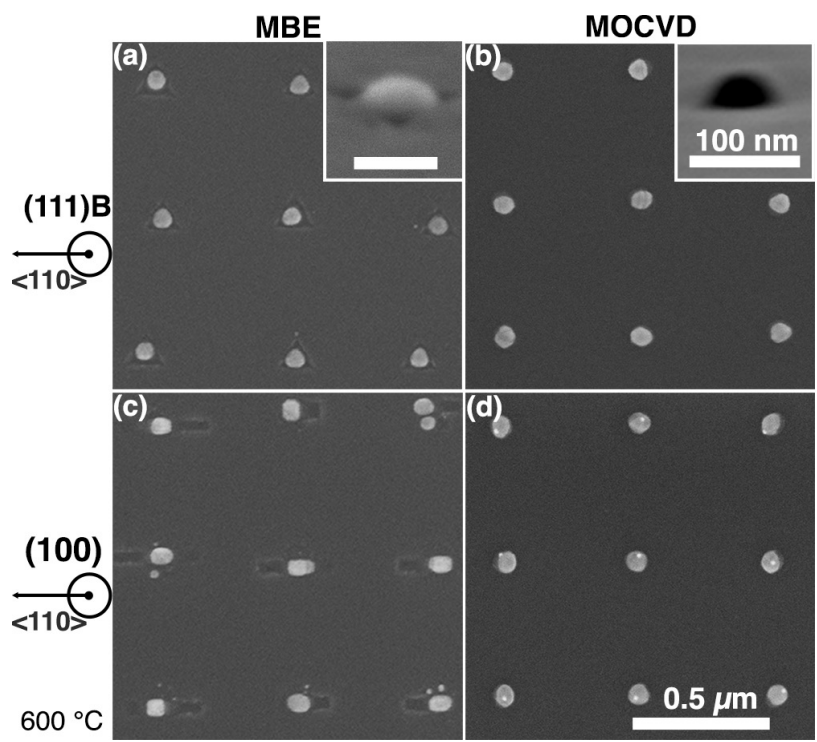

**Figure 8.** SEM images of MBE (a,c) and MOCVD (b,d) annealing at 600 °C of AuAg NP arrays. NP diameter is 100nm and pitch distance is 0.5μm. GaAs (111)B and (100) surface orientation is shown in the top and bottom images, respectively. The insets are 85° tilted SEM images, which illustrates the characteristics of the NPs at these annealing conditions. (All primary images share the same 0.5 μm scale bar and the insets have a 100 nm scale bar.)

ASSOCIATED CONTENT

**Supporting Information**. A Supplemental Information file is associated with this work. This include Energy-dispersive X-ray spectroscopy data, Diameter dependence of planar nanowires, Comparison between Au and Ag nanoparticle splitting, Volatility of Ag nanoparticles, and Diameter effects for Au-Ag nanoparticles. This material is available free of charge via the Internet at http://pubs.acs.org.

AUTHOR INFORMATION

**Corresponding Author**

* jessica.bolinsson@nbi.ku.dk, kimberly.dick@ftf.lth.se

Author Contributions

The manuscript was written through contributions of all authors. All authors have given approval to the final version of the manuscript.

ACKNOWLEDGMENT

J.B. gratefully acknowledges financial support from the Villum Foundation (Young Investigator Program, no. 10121). J.N. and J.B. also acknowledges support of the ANaCell project of the Strategic Research Council (Innovation Fund Denmark), the KU2016 Excellence program BioSynergy. K.A.D and E.K.M. acknowledges financial support from the European Research Council under the European Unions Seventh Framework Programme (FP/2007-2013)/ERC Grant Agreement No. 336126 and the Knut and Alice Wallenberg Foundation (KAW). Further, we thank Jonas Johansson for fruitful discussions and for excellent technical assistance we thank Claus B. Sørensen, Thomas Kanne and Caroline Lindberg. The Center for Quantum Devices is funded by the Danish Natural Research Foundation (DG).

Insert Table of Contents Graphic and Synopsis Here

*Supplemental Information*
*to*

Annealing of Au, Ag and Au-Ag alloy nanoparticle arrays

on GaAs (100) and (111)B

*Alexander M. Whiticar[1], Erik Mårtensson[2], Jesper Nygård[1], Kimberly A. Dick[2,3*] and Jessica Bolinsson[1*]*

[1] Center for Quantum Devices & Nano-Science Center, Niels Bohr Institute,

University of Copenhagen, Nørregade 10, 1165 København, Denmark.

[2] Solid State Physics/NanoLund, [3] Center for Analysis and Synthesis, Lund University,

221 00 Lund, Sweden.

* Corresponding author: jessica.bolinsson@nbi.ku.dk , kimberly.dick@ftf.lth.se

# I. Energy-dispersive X-ray spectroscopy of point

The data shown in Fig S1(b) is acquire from Energy-dispersive X-ray spectroscopy (EDS) detector point scans on substrates that have been cleaved to bisect the nanoparticle (NP) arrays. The cleaved substrates are analyzed by scanning electron microscope (SEM) equipped with an EDS detector where they are mounted on a cross-sectional sample holder. Point scans of the substrate and NP are acquired at 5 locations (1-5 in Fig. S1 (a)) chosen to investigate the change in Ga to As composition in the substrate and NP.

The data shown in Fig. S1(b) shows the change in Ga:As composition. The raw data acquired from the EDS detector had the background bremsstrahlung radiation subtracted using Kramer's law. The area enclosed by the Ga and As peaks was calculated to allow for compositional comparison using the Cliff-Lorimer method. Location 1 was assumed to have equal Ga and As stoichiometry and used as a standard for the four other locations.

At 400 °C, there is a decrease in the As molar fraction as the scan location nears the NP. This data suggest that there is a higher Ga content in or around the droplet, suggesting that an alloy between Au and Ga has formed. This depression in the As molar fraction is even more pronounced at 600 °C, where the intermediary substrate location (2) shows a much larger decrease in As.

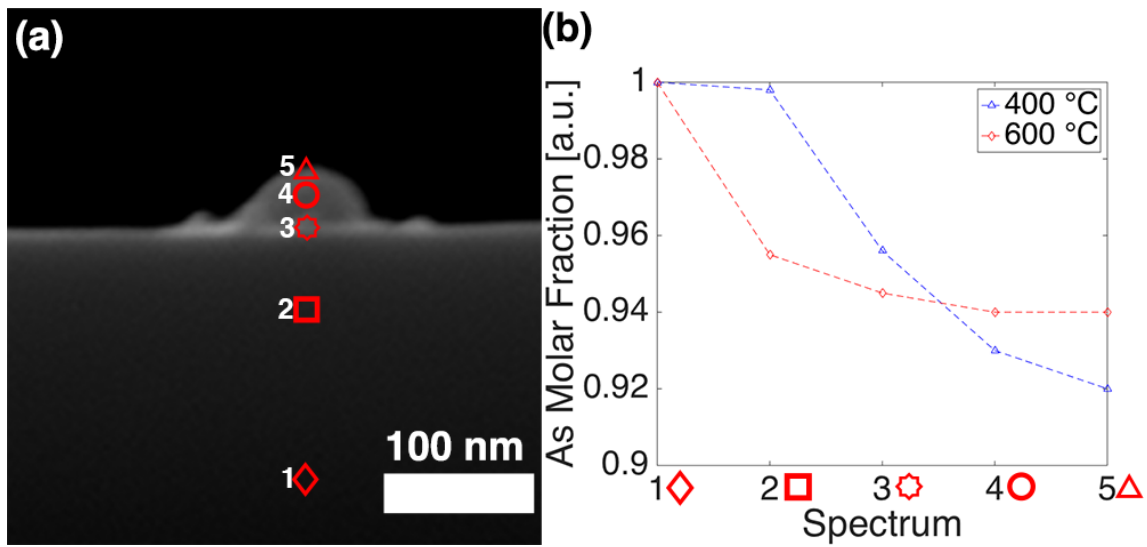

**S1:** (a) Cross-sectional scanning electron micrograph of a Au NP on a GaAs substrate with the characteristic locations where EDS point spectra were obtained. The numbers correspond to the spectrum locations in (b). (b) Calculated Cliff — Lorimer factors for the scan locations in (a) for 400 and 600 °C annealing.

## II. Diameter dependence of planar nanowires

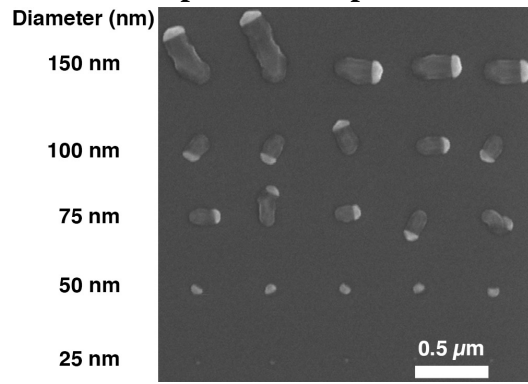

**S2:** SEM images of a Au NP array with increasing NP diameters annealed at 650 °C by a MOCVD system. The planar NWs length and width is dependent on the size of the initial NP.

## III. Comparison between Au and Ag nanoparticle splitting

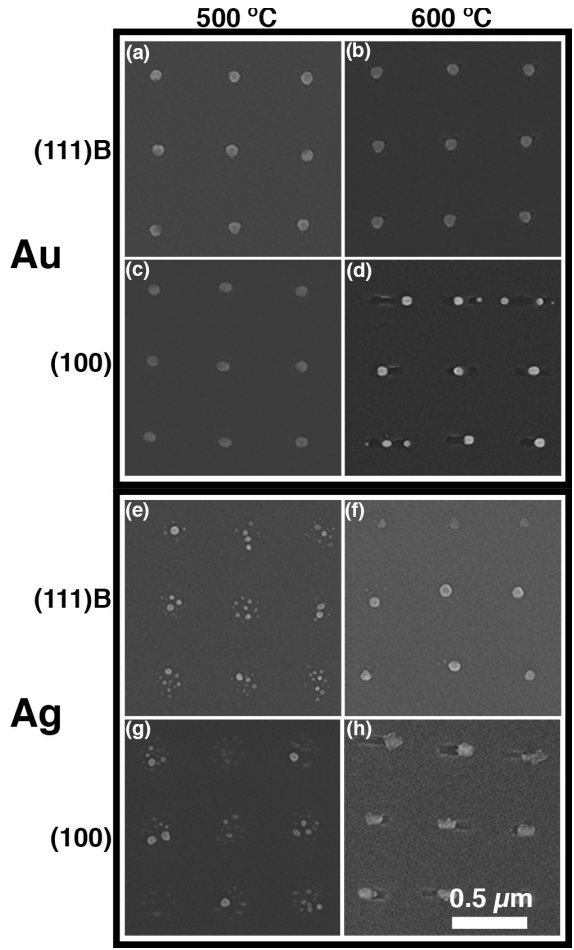

**S3:** SEM images of Ag and Au NPs annealed by MBE on (111)B and (100) substrates to compare the different types of splitting and mobility. The splitting of Ag NPs is substrate orientation independent, while for Au it only occurs on the (100) orientation for specific diameters and pitch distances (see Fig. 3 of the main text). All arrays consist of 100 nm NPs with a 0.5 µm pitch distance, expect for (d), which have 50 nm NPs to show the Au splitting behaviour.

## IV. Volatility of Ag NPs

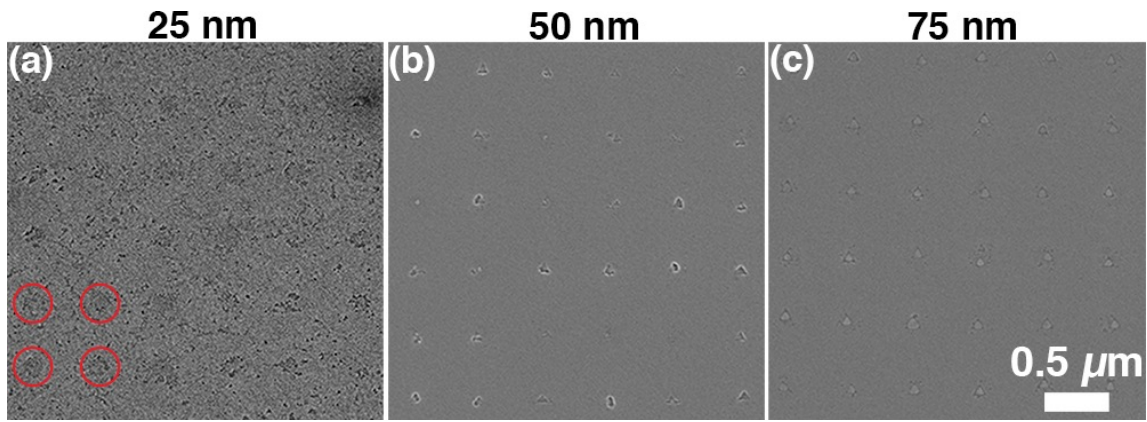

**S4:** Scanning electron micrographs of Ag NPs on GaAs (111)B substrates annealed at 600 °C by a MBE system. (a) The 25 nm Ag NPs have disappeared from the substrate surface. The red circles indicate the Ag NPs position before annealing, with noticeable changes to the substrate. The 50 (b) and 75 (c) nm NPs etch holes into the substrate where they remain after annealing. All arrays have a 0.5 $\mu$m pitch distance. Scale bar is common to all images

## V. Diameter Effects for Au-Ag NPs

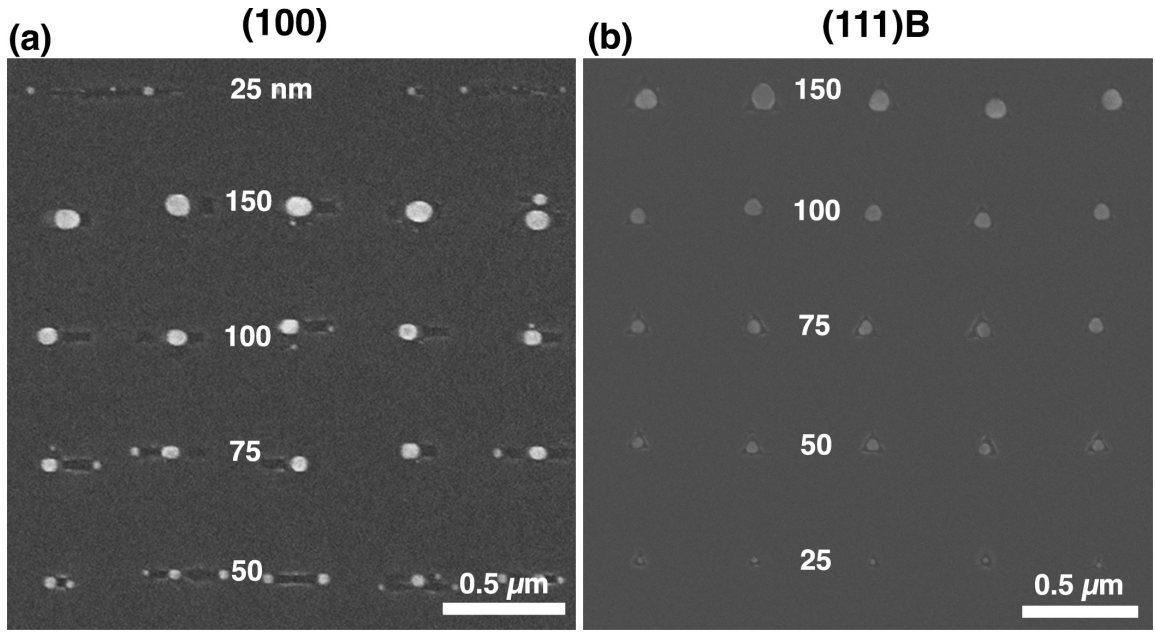

**S5:** The diameter dependence of Au-Ag NPs annealed by a MBE system at 600 °C. (a,b) Au-Ag arrays with diameters varying from 25 to 150 nm in vertical increments as indicated.